\documentclass[twocolumn,showpacs,preprintnumbers,nofootinbib]{revtex4}
\usepackage{epsfig,amsmath,amssymb,color}

\newcommand{\beq}{\begin{equation}}
\newcommand{\eeq}{\end{equation}}
\newcommand{\bea}{\begin{eqnarray}}
\newcommand{\eea}{\end{eqnarray}}
\newcommand{\Fig}[1]{Fig.\,\ref{#1}}

\newcommand{\Eq}[1]{Eq.\,(\ref{#1})}

\newcommand{\Eqsand}[2]{Eqs.\,(\ref{#1}) and (\ref{#2})}
\newcommand{\Eqsto}[2]{Eqs.\,(\ref{#1}) to (\ref{#2})}

\newcommand{\Sec}[1]{Sec.\,\ref{#1}}

\newcommand{\f}{\frac}
\newcommand{\non}{\nonumber}

\newcommand{\gp}{g'}
\newcommand{\as}{\alpha_s}
\newcommand{\aem}{\alpha_{\rm em}}
\newcommand{\asMSbar}{\alpha^{\overline{\rm MS}}_s}
\newcommand{\aemMSbar}{\alpha^{\overline{\rm MS}}_{\rm em}}
\newcommand{\sw}{s_W}

\newcommand{\sws}{s^2_W}

\newcommand{\MW}{M_W}
\newcommand{\MZ}{M_Z}

\newcommand{\mh}{M_{h^0}}
\newcommand{\MH}{M_{H^0}}
\newcommand{\MA}{M_{A^0}}

\newcommand{\mb}{m_b}
\newcommand{\mt}{m_t}

\newcommand{\mmu}{m_\mu}

\newcommand{\mgl}{M_{\tilde g}}

\newcommand{\Heff}{{\cal H}_{\rm eff}}
\newcommand{\Leff}{{\cal L}_{\rm eff}}

\newcommand{\hc}{{\rm h.c.}}
\newcommand{\GF}{G_F}
\newcommand{\BR}{{\cal B}}
\newcommand{\MeV}{{\rm MeV}}
\newcommand{\GeV}{{\rm GeV}}
\newcommand{\TeV}{{\rm TeV}}
\newcommand{\MSbar}{\overline{\rm MS}}
\newcommand{\DRbar}{\overline{\rm DR}}

\newcommand{\ord}{{\cal O}}
\newcommand{\eps}{\epsilon}

\newcommand{\sL}{{\scalebox{0.6}{$L$}}}
\newcommand{\sR}{{\scalebox{0.6}{$R$}}}

\newcommand{\BXsga}{\bar{B} \to X_s \gamma}

\newcommand{\BXsll}{\bar{B} \to X_s l^+ l^-}

\newcommand{\Bdmm}{B_d \to \mu^+ \mu^-}
\newcommand{\Bsmm}{B_s \to \mu^+ \mu^-}
\newcommand{\Bdsmm}{B_{d,s} \to \mu^+ \mu^-}

\newcommand{\BRBd}{{\cal B} (\Bdmm)}
\newcommand{\BRBs}{{\cal B} (\Bsmm)}
\newcommand{\BRBds}{{\cal B} (\Bdsmm)}
\newcommand{\BRga}{{\cal B} (\BXsga)}
\newcommand{\BRll}{{\cal B} (\BXsll)}

\newcommand{\btosgamma}{b \to s \gamma}

\newcommand{\CP}{C\hspace{-0.25mm}P}

\newcommand{\mysigma}{\hspace{0.4mm} \sigma}
\newcommand{\TB}{t_\beta}

\newcommand{\SB}{s_\beta}
\newcommand{\CB}{c_\beta}
\newcommand{\SA}{s_\alpha}
\newcommand{\CA}{c_\alpha}

\newcommand{\mstl}{m_{\tilde{t}_L}}
\newcommand{\mstr}{m_{\tilde{t}_R}}

\newcommand{\msbr}{m_{\tilde{b}_R}}

\newcommand{\mstaur}{m_{\tilde{\tau}_R}}

\newcommand{\etal}{{\it et al}.}
\newcommand{\epszero}{\epsilon_0}
\newcommand{\epsthree}{\tilde{\epsilon}_3}
\newcommand{\epsY}{\epsilon_Y}

\newcommand{\epsGP}{\epsilon_{\rm GP}}
\newcommand{\deltaGP}{\delta_{\rm GP}}

\newcommand{\mbbar}{\overline{m}_b}
\newcommand{\msbar}{\overline{m}_s}
\newcommand{\mdbar}{\overline{m}_d}
\newcommand{\mdibar}{\overline{m}_{d_i}}
\newcommand{\mtbar}{\overline{m}_t}
\newcommand{\mfbar}{\overline{m}_f}
\newcommand{\mlbar}{\overline{m}_{\tau}}
\newcommand{\BdsBbards}{B_{d,s}\text{--}\bar{B}_{d,s}}
\newcommand{\BdBbard}{B_d\text{--}\bar{B}_d}
\newcommand{\MSUSY}{M_{\rm SUSY}}
\newcommand{\DMK}{\Delta M_K}
\newcommand{\epsK}{|\epsilon_K|}
\newcommand{\DMd}{\Delta M_d}
\newcommand{\DMs}{\Delta M_s}
\newcommand{\DMds}{\Delta M_{d,s}}
\newcommand{\Btn}{B^+ \to \tau^+ \nu_{\tau}}
\newcommand{\BRBtn}{{\cal B} (B^+ \to \tau^+ \nu_{\tau})}
\newcommand{\tb}{\tan \beta}
\newcommand{\CAB}{c_{\alpha - \beta}}
\newcommand{\SAB}{s_{\alpha - \beta}}
\newcommand{\CBA}{c_{\beta - \alpha}}
\newcommand{\SBA}{s_{\beta - \alpha}}

\begin{document}

\preprint{ZU-TH 1/07} 

\title{
\boldmath
Supersymmetric large $\tb$ corrections to $\DMds$ and $\Bdsmm$ revisited
\unboldmath
}

\author{Ayres~Freitas, Esther~Gasser and Ulrich~Haisch} 

\affiliation{
Institut f\"ur Theoretische Physik, Universit\"at Z\"urich,
CH-8057 Z\"urich, Switzerland 
}

\date{\today}

\begin{abstract}
\noindent
We point out that in the minimal supersymmetric standard model terms
from the mixing of Higgs and Goldstone bosons which are connected to
the renormalization of $\tb$ via Slavnov-Taylor identities give rise
to corrections that do not vanish in the limit where the
supersymmetric particles are much heavier than the Higgs bosons. These
additional contributions have important phenomenological implications
as they can lead to potentially large supersymmetric effects in $\DMd$
and to a significant increase of $\DMs$ relative to the standard model
prediction for a light pseudoscalar Higgs $A^0$. 
We calculate all the missing one-loop pieces and combine
them with the known effective non-holomorphic terms to obtain improved
predictions for the $\BdsBbards$ mass differences $\DMds$ and the
branching ratios of $\Bdsmm$ in the large $\tb$ regime of the minimal
supersymmetric standard model with minimal-flavor-violation.

\end{abstract}


\maketitle

\section{Introduction}
\label{sec:introduction}

In minimal supersymmetric (SUSY) extensions of the standard model (SM)
soft SUSY breaking terms are introduced that explicitly violate the
underlying symmetry without spoiling the cancellation of quadratically
divergent radiative corrections to the Higgs and other scalar masses.
These soft terms must have positive mass dimension and the scale
$\MSUSY$ associated with them should be below a few TeV to naturally
maintain the hierarchy between the electroweak scale $v$ and the
Planck or any other very large energy scale. While theoretically
little is known definitely about the origin and mechanism of the SUSY
breaking itself, the soft mass terms will be measured and constrained
as superpartners are detected. If SUSY is the solution to the
hierarchy problem, then the Tevatron may, and the LHC will likely, find
direct evidence for it. Meanwhile, the soft terms are also indirectly
constrained by low-energy observables such as $\DMds$, $\BRga$,
$\BRBds$, $\BRBtn$, and $(g - 2)_\mu$. From these measurements one can
learn about the structure of SUSY breaking.
  
For large sparticle masses, i.e., $v/\MSUSY \to 0$, the effects of
SUSY degrees of freedom can be absorbed into the coupling constants of
local operators in an effective theory that arises after decoupling
the heavy particles. The corresponding low-energy theory is a
two-Higgs-doublet model (THDM) of type II. In order for the THDM to be
well-defined beyond tree-level, the effective couplings need to be
calculated in the limit of unbroken $SU(2) \times U(1)$ symmetry. At
the matching scale, some of the corrections to the effective couplings
do not vanish for $\MSUSY \to \infty$ if the Higgsino parameter $\mu$
is assumed to be of comparable size, i.e., $\mu = \ord (\MSUSY)$. In
addition, some of the corrections can be enhanced by the ratio $\tb =
v_u/v_d$ of the vacuum expectation values (VEVs) $v_{d, u}$ of the two
Higgs doublets $H_{d, u}$ that separately give masses to the down- and
up-type fermions. As a result, they can be sizable, of order $\as \tan
\beta \simeq 1$ for values of $\tb \gg 1$, and need to be resummed if
applicable.

In the minimal supersymmetric SM (MSSM) with minimal-flavor-violation
(MFV), four different types of large $\tb$ contributions that affect
the interactions between Higgs bosons and SM fermions have been
identified: (i) corrections to the vertices between a Higgs boson and
down-type fermions \cite{Hall:1993gn}, which are interpreted as
corrections to the Yukawa couplings $y_d$ and resummed
\cite{Carena:1999py}, (ii) similar corrections to the vertices between
a Higgs boson and up-type fermions \cite{ltbup} which are not
resummed, (iii) corrections to the Cabibbo-Kobayashi-Maskawa (CKM)
matrix elements \cite{Blazek:1995nv}, and (iv) flavor-changing neutral
Higgs vertex corrections \cite{Babu:1999hn}, which do not appear at
tree-level.

The purpose of this article is it to point out that there are
additional terms from mixing of Higgs and Goldstone bosons \cite{UN}
which are connected to the renormalization of $\tb$. These new
contributions have important phenomenological implications as they can
lead to potentially large SUSY effects in $\DMd$ and to a significant
increase of $\DMs$ relative to the SM prediction in a certain region
of the allowed parameter space. Both findings go against common lore
\cite{BCRS, Isidori:2001fv, D'Ambrosio:2002ex, Carena:2006ai,
  Blanke:2006ig, recent}, but they are an unavoidable consequence of
the analysis presented here.

This article is organized as follows. In the next section we derive
the $\tb$ enhanced corrections to the effective Higgs interactions
with quarks of the third generation, including all terms that arise
from the one-loop mixing between eigenstates of the two Higgs
doublets. Analytic formulas for the neutral Higgs contributions to
$\DMds$ and $\BRBds$ are presented in \Sec{sec:DP}. \Sec{sec:numerics}
contains a numerical analysis of $\DMds$ in the MFV MSSM with large
$\tb$ taking into account all relevant constraints from flavor and
collider physics. Concluding remarks are given in
\Sec{sec:conclusions}.

\section{Effective theory}
\label{sec:effectivetheory}

Beyond leading order, the physical Higgs fields $A^0$ and $H^\pm$ can
mix with the longitudinal components of the gauge bosons fields $Z^0$
and $W^\pm$ through loop corrections. This mixing has to be removed
for on-shell momenta by suitable rotations of the fields. Due to the
connection between the longitudinal gauge bosons and the Goldstone
bosons $G^0$ and $G^\pm$, this procedure also implies field
renormalization for the mixing between Higgs and Goldstone bosons. The
relation between the terms with gauge and Goldstone bosons is given by
Slavnov-Taylor identities (STIs), derived from the invariance of
two-point functions under Becchi-Rouet-Stora-Tyutin
transformations. In our case the relevant STIs are given to one-loop
order by
\beq \label{eq:STIs}
0 = k^\mu \hat{\Sigma}_\mu^{V \hspace{-0.25mm} S} (k) +
\widetilde{M}_V \hat{\Sigma}^{S^\prime \hspace{-0.25mm} S} (k^2) +
\ldots \, ,
\eeq 
where $\{V, S, S^\prime, \widetilde{M}_V\} = \{Z^0, A^0, G^0, i
\MZ\}$, $\{W^\pm, H^\mp,$ $G^\pm, \pm \MW\}$, and hatted quantities
denote amputated and renormalized mixing self-energies. The ellipses
stand for terms that vanish in the limit $k^2 \to M_S^2$ which are
irrelevant for the further discussion.

In the renormalized mixing self-energies, besides the field
renormalization counterterms $\delta Z^{S^\prime \hspace{-0.25mm} S}$,
the contributions from the tadpole renormalization, $\delta t_{d,u}$,
and from the renormalization of the VEVs, $\delta Z_{d,u}$, need to be
included. The tadpole counterterms are fixed by the requirement that
the properly minimized scalar potential should have no finite tadpole
terms, $\hat{t}_{d, u} = t_{d, u} + \delta t_{d, u} = 0$. The
renormalization of the VEVs, $\hat{v}_{d, u} = (1 + \delta Z_{d, u})
v_{d, u}$, translates into the renormalization of the gauge boson
masses and $\tb$, $\delta \tb = (\delta Z_u - \delta Z_d) \tb$. Here
quantities without hat denote unrenormalized contributions.

\begin{figure}[!t]
\vspace{1mm}
\scalebox{0.6}{\includegraphics{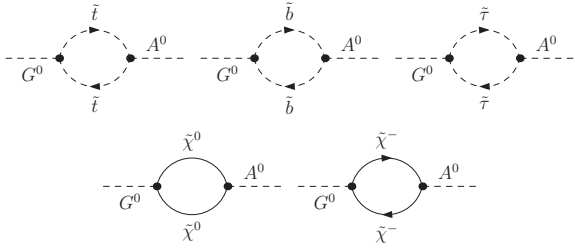}}
\vspace{-1mm}
\caption{Diagrams in the MSSM that lead to non-decoupling corrections
  to the $G^0 A^0$ mixing self-energy. See text for details.}
\label{fig:g0a0}
\end{figure}

Decomposing the $V \hspace{-0.25mm} S$ mixing self-energy via 
$\hat{\Sigma}_\mu^{V \hspace{-0.25mm} S} (k) = k_\mu \hat{\Sigma}^{V
\hspace{-0.25mm} S} (k^2)$ and using the above conventions the
renormalized self-energies can be written as   
\beq \label{eq:Sigmahats}
\begin{split}
  \hat{\Sigma}^{V \hspace{-0.52mm} S} (k^2) & = \Sigma^{V
    \hspace{-0.25mm} S} (k^2) - \widetilde{M}_V \f{1}{2} \delta
  Z^{S^\prime \hspace{-0.25mm} S} + \widetilde{M}_V \CB^2 \, \delta
  \TB
  \, , \\
  \hat{\Sigma}^{S^\prime \hspace{-0.25mm} S} (k^2) & =
  \Sigma^{S^\prime \hspace{-0.25mm} S} (k^2) + \frac{1}{2} k^2 \delta
  Z^{S^\prime
    \hspace{-0.25mm} S} - M_S^2 \CB^2 \, \delta \TB \\
  & + \f{g}{2 \MW} \left ( \CBA \, \delta t_{h^0} - \SBA \, \delta
    t_{H^0} \right ) + \ldots \, ,
\end{split}
\eeq 
where again terms that vanish on-shell have been omitted. Furthermore
the abbreviations $\SB = \sin \beta$, $\CB = \cos \beta$, $\TB = \tb$,
etc. have been used and the tadpoles of the $H_{d, u}$ fields
expressed in terms of those of the Higgs mass eigenstates $h^0$ and
$H^0$.

Requiring now that the mixing between longitudinal gauge and Higgs
bosons vanishes when the Higgs boson momentum is on-shell,
$\hat{\Sigma}^{V \hspace{-0.25mm} S} (M_S^2) = 0$, one obtains for the
field renormalization counterterms  
\beq \label{eq:dZSS}
\begin{split}
  \delta Z^{S^\prime \hspace{-0.25mm} S} & = \f{2}{ \widetilde{M}_V}
  \Sigma^{V \hspace{-0.25mm} S} (M_S^2) + 2 \CB^2 \, \delta \TB \\
  & = -\f{2}{M_S^2} \Sigma^{S^\prime \hspace{-0.25mm} S} (M_S^2) + 2
  \CB^2 \, \delta \TB  \\
  & - \f{g}{M_S^2 \MW} \left ( \CBA \, \delta t_{h^0} - \SBA \, \delta
    t_{H^0} \right ) \, .
\end{split}
\eeq 
The presence of $\delta \TB$ in the above equalities illustrates that
beyond tree-level a definition of $\tb$ is needed, which in turn
controls the extent of mixing between Higgs and Goldstone boson
fields.

Several prescriptions for the renormalization of $\tb$ have been
studied in the literature, but from a careful analysis
\cite{Freitas:2002um} it was found that in order to avoid large higher
order corrections, the best scheme is to use $\DRbar$ renormalization
for $\tb$, which is manifestly process-independent to all orders and
gauge-independent at the one-loop level within the class of $R_{\xi}$
gauges. If only the leading contributions for large $\tan \beta$ are
retained, the counterterm $\delta \tb$ is thus identical to zero.  
  
In the limit $\tb \gg 1$, the leading contribution to the field
renormalization counterterms involving SUSY loops are of zeroth order
in $\tb$. The Feynman diagrams that lead to the non-decoupling
corrections to the $G^0 A^0$ mixing self-energy are shown in
\Fig{fig:g0a0}. Calculating these, the corresponding $G^\pm H^\mp$
diagrams, and the tadpole counterterms in the limit of unbroken $SU(2)
\times U(1)$ symmetry we find   
\beq \label{eq:ZGAZGH}
\f{1}{2} \delta Z^{G^0 A^0} = \f{1}{2} \delta Z^{G^\pm H^\mp} =
\eps_{\rm GP} \, ,
\eeq
where\footnote{Our sign convention for the tri-linear soft
  SUSY breaking couplings $A_f$ is fixed by the left-right sfermion
  mixing which in the case of the stop reads $\mt (A_t - \mu \cot
  \beta)$.} 
\begin{align} \label{eq:epsGP}
  \epsGP & = -\frac{\mu}{32 \pi^2} \Bigg [ 3 y_t^2 \f{A_t}{\mstr^2}
  H_1 (x_{\tilde{t}}) + 3 y_b^2 \f{A_b}{\msbr^2} H_1 (x_{\tilde{b}}) \\
  & + y_\tau^2 \f{A_\tau}{\mstaur^2} H_1 (x_{\tilde{\tau}}) + \gp^2
  \f{M_1}{\mu^2} H_1 (x_1) + 3 g^2 \f{M_2}{\mu^2} H_1 (x_2) \Bigg ] ,
  \non
\end{align}
and 
\beq \label{eq:H1} 
H_1 (x) = \f{x + 1}{(x - 1)^2} - \f{2 x}{(x - 1)^3} \ln x \, .
\eeq
The variables $x_{\tilde{f}} = m^2_{\tilde{f}_L}/m^2_{\tilde{f}_R}$
and $x_i = M^2_i/\mu^2$ denote the ratios of the left- and
right-handed sfermion masses and the soft SUSY breaking masses
$M_{1,2}$ and the Higgsino parameter $\mu$ squared. We assume $\CP$
conservation, so all soft SUSY breaking terms are real. 

For sizable values of the tri-linear soft SUSY breaking couplings
$A_{t,b}$ the correction $\epsGP$ is dominated by the stop and sbottom
contributions, which are proportional to the square of the Yukawa
couplings $y_{t,b}$.  The stau and bino corrections are numerically
insignificant. Assuming the values of $|\mu|$, $|A_t|$, $\mstl$, and
$\mstr$ to be degenerate and keeping only the stop contribution one
finds $\epsGP \simeq -{\rm sign} (\mu A_t)/(32 \pi^2) \simeq -3 \times
10^{-3} \, {\rm sign} (\mu A_t)$, while $| \epsGP |$ can reach $\simeq
10^{-2}$ for $\tb \gg 1$ and natural choices of sfermion masses and
soft SUSY breaking parameters in the $\TeV$ range.

The corrections \eqref{eq:ZGAZGH} as such are not $\tb$
enhanced. However, since they describe the mixing between eigenstates
of the two Higgs doublets, they replace $\tb$ suppressed Higgs-fermion
couplings with the non-suppressed Goldstone-fermion couplings. The
$\tb$ suppression at tree-level is therefore effectively lifted at the
one-loop level. Obviously, this can only occur at the next-to-leading
order (NLO) and there are no resummable enhanced $\tb$ corrections
beyond that order.

The term $\epsGP$ can also be derived from the STIs of the
$\CP$-even neutral Higgs sector. In this case one
demands un-mixing of the on-shell Higgs bosons at loop level, and
needs to take into account the renormalization of other parameters of
the Higgs potential in the broken phase, such as the on-shell
renormalization of the gauge boson masses. We find
\beq \label{eq:dZhHZHh}
\f{\mh^2 - \MH^2}{2 \MH^2} \delta Z^{h^0 H^0} = \f{\MH^2 - \mh^2}{2
  \mh^2} \delta Z^{H^0 h^0} = \epsGP \, .
\eeq       
As a result, the mixing of $h^0$ and $H^0$ receives particularly large
contributions from $\epsGP$ if the difference between the masses $\mh$
and $\MH$ is small.

The term $\epsGP$ essentially describes the mixing between the two
Higgs doublets at one loop. Therefore the resulting contributions are
universal for the different $\CP$ and charge scalar eigenstates, up to
the different mixing in the $\CP$-even neutral sector, which is
described here through the Higgs masses in \Eq{eq:dZhHZHh}.

The diagonalization of the Higgs mass matrix necessarily induces
diagonal and off-diagonal couplings of the neutral and charged scalar
to the quark fields. In the basis of neutral and charged Higgs mass
eigenstates the large $\tb$ corrections to the effective Higgs
interactions with quarks of the third generation can be cast into the
following form
\begin{widetext}
\begin{align} \label{eq:lagrangian} 
  \Leff & = \GF^{1/2} 2^{1/4} \Bigg \{ \f{\mbbar\TB}{1 + \epsthree \,
    \TB} \Bigg [ i \bar{b}_{\sL} b_{\sR} A^0 + \bigg ( \frac{\SA}{\SB}
  - \frac{\CA}{\SB} \, \epsthree + \f{\mh^2}{\MH^2 -
    \mh^2} \frac{\CA}{\SB} \, \epsGP \bigg ) \bar{b}_{\sL} b_{\sR} h^0  \non \\
  & \hspace{10em} - \bigg ( \frac{\CA}{\SB} + \frac{\SA}{\SB} \,
  \epsthree \bigg ) \bar{b}_{\sL} b_{\sR} H^0 + \sqrt{2} V_{tb}^{\rm
    eff} \, \bar{t}_{\sL} b_{\sR} H^+
  \Bigg ] \non \\
  & + \frac{\epsY y_t^2 \TB^2}{(1 + \epsthree \, \TB) (1 + \epszero \,
    \TB)} \Bigg [ i \left ( V_{tb}^{{\rm eff} \ast} V_{t d^i}^{\rm
      eff} \, \mbbar \, \bar{b}_{\sR} d^i_{\sL} A^0 - V_{td^i}^{{\rm
        eff} \ast} V_{t b}^{\rm eff} \, \mdibar \, \bar{b}_{\sL}
    d^i_{\sR} A^0 \right ) \non \\ \non \\[-10mm] \\[0mm] 
  & \hspace{10em} - \bigg ( \frac{\CAB}{\SB^2} + \f{\mh^2}{\MH^2 -
    \mh^2} \frac{\CA}{\SB} \, \epsGP \bigg ) \left ( V_{tb}^{{\rm eff}
      \ast} V_{t d^i}^{\rm eff} \, \mbbar \, \bar{b}_{\sR} d^i_{\sL}
    h^0 + V_{td^i}^{{\rm eff} \ast} V_{t b}^{\rm eff} \, \mdibar \,
    \bar{b}_{\sL} d^i_{\sR} h^0 \right )
  \non \\
  & \hspace{10em} - \frac{\SAB}{\SB^2} \left ( V_{tb}^{{\rm eff} \ast}
    V_{t d^i}^{\rm eff} \, \mbbar \, \bar{b}_{\sR} d^i_{\sL} H^0 +
    V_{td^i}^{{\rm eff} \ast} V_{t b}^{\rm eff} \, \mdibar \,
    \bar{b}_{\sL} d^i_{\sR} H^0 \right ) \Bigg ]
  \non \\
  & + \mtbar \Bigg [ \big ( \TB^{-1} - \epszero' - \epsY' y_b^2 +
  \epsGP \big ) \big ( i \bar{t}_{\sL} t_{\sR} A^0 + \sqrt{2}
  V_{tb}^{\rm eff} \, \bar{t}_{\sR} b_{\sL} H^+ \big ) - \f{\CA}{\SB}
  \bar{t}_{\sL} t_{\sR} h^0 - \f{\SA}{\SB} \bar{t}_{\sL} t_{\sR} H^0
  \Bigg ] \Bigg \} + \hc \, , \non 
\end{align}
\end{widetext}
for $d^i = d, s$. Here $\GF$ denotes the Fermi constant, $V_{ij}^{\rm
  eff}$ are the physical CKM matrix elements, and $\mfbar$ are running
$\MSbar$ masses evaluated at a scale of order $\mt$, which are
connected to the Yukawa couplings through $y^2_t = 2 \sqrt{2} \GF
\mtbar^2$ and $y^2_b = 2 \sqrt{2} \GF \mbbar^2 \TB^2/(1 + \epsthree \,
\TB)^2$.  The subscripts $L$ and $R$ indicate the chirality of the
quark fields involved in the interaction.

As the field renormalization counterterms $\delta Z^{S^\prime S}$ in
\Eqsand{eq:ZGAZGH}{eq:dZhHZHh} cancel the momentum-independent part of
the mixing self-energies $\Sigma^{S^\prime S}(k^2)$, dimension four
operators related to the mixing of the scalar fields $\{S, S^\prime\}
= \{A^0, G^0\}$, $\{H^\mp, G^\pm\}$, $\{h^0, H^0\}$, and $\{H^0,
h^0\}$ are removed from the effective theory. Alternatively, if the
vanishing of the $S^\prime S$ mixing in the full theory is not
enforced by suitable renormalization conditions, then diagrams with
insertions of dimension four operators that mix the scalar fields
$S^\prime$ and $S$ will contribute on the effective side. If
implemented correctly, the two strategies lead naturally to identical
results for physical observables.

The epsilon parameters $\epsthree$, $\epszero$, $\epsY$, $\epszero'$,
and $\epsY'$ are defined as in \cite{BCRS}. In our numerical analysis
we employ them in the limit of unbroken $SU(2) \times U(1)$ and
include all effects proportional to the $SU(2)$ couplings $g$ and
$\gp$ squared. The corresponding analytic expressions read
\begin{align} \label{eq:epstilde3}
  \epsthree &= \epszero + \epsY y_t^2 \, ,
  \displaybreak[0] \\[2mm]
  \epszero &= -\f{2 \as}{3 \pi} \f{\mu}{\mgl} H_2 (u_{\tilde{t}_L},
  u_{\tilde{b}_R}) \non \displaybreak[0] \\ & +\f{1}{16 \pi^2} \bigg [
  \f{\gp^2}{6} \f{M_1}{\mu} \big ( H_2 (v_{\tilde{t}_L}, x_1) + 2
  H_2 (v_{\tilde{b}_R}, x_1) \big ) \displaybreak[0] \\
  & \hspace{2.5em} + \f{\gp^2}{9} \f{\mu}{M_1} H_2 ( w_{\tilde{t}_L},
  w_{\tilde{b}_R}) + \f{g^2}{2} \f{M_2}{\mu} H_2 (v_{\tilde{t}_L},
  x_2) \bigg ] , \non
  \displaybreak[0] \\[2mm]
  \epsY &= \f{-1}{16 \pi^2} \bigg [ \f{A_t}{\mu} H_2 (v_{\tilde{t}_L},
  v_{\tilde{t}_R}) - \f{g^2}{y_t^2} \f{M_2}{\mu} H_2 (v_{\tilde{t}_L},
  x_2) \bigg ] ,
  \displaybreak[0] \\[2mm]
  \epszero' &= -\f{2 \as}{3 \pi} \f{\mu}{\mgl} H_2 (u_{\tilde{t}_L},
  u_{\tilde{t}_R}) \non \\ &- \f{1}{16 \pi^2} \bigg [ \f{\gp^2}{6}
  \f{M_1}{\mu} \big ( H_2 (v_{\tilde{t}_L}, x_1) - 4
  H_2 (v_{\tilde{t}_R}, x_1) \big ) \\
  & \hspace{2.5em} + \f{2\gp^2}{9} \f{\mu}{M_1} H_2 ( w_{\tilde{t}_L},
  w_{\tilde{t}_R}) - \f{g^2}{2} \f{M_2}{\mu} H_2 (v_{\tilde{t}_L},
  x_2) \bigg ] , \non \displaybreak[0] \\[2mm] \label{eq:epsYprime}
  \epsY' &= \f{-1}{16 \pi^2} \bigg [ \f{A_b}{\mu} H_2
  (v_{\tilde{t}_L}, v_{\tilde{b}_R}) - \f{g^2}{y_b^2} \f{M_2}{\mu} H_2
  (v_{\tilde{t}_L}, x_2) \bigg ] .
\end{align}
Here $u_{\tilde{q}} = m^2_{\tilde{q}}/\mgl^2$, $v_{\tilde{f}} =
m^2_{\tilde{f}}/\mu^2$, $w_{\tilde{f}} = m^2_{\tilde{f}}/M_1^2$, and
\beq \label{eq:H2} 
H_2 (x, y) = \f{x \ln x}{(1 - x) (x - y)} + \f{y \ln y}{(1 - y) (y -
  x)} \, .
\eeq 
Our analytic results for $\epsthree$, $\epszero$, $\epsY$,
$\epszero^\prime$, and $\epsY^\prime$ have been obtained in the
approximation $V_{ud} \simeq V_{cs} \simeq V_{tb} \simeq 1$. They
agree with the corresponding expressions in \cite{Hall:1993gn,
  Carena:1999py, Blazek:1995nv, Babu:1999hn, ltbup, BCRS,
  Isidori:2001fv, D'Ambrosio:2002ex} for $g = \gp = 0$.

\section{\boldmath Double penguin contributions to $\DMds$ and $\BRBds$} 
\label{sec:DP}

The unique role of neutral Higgs double penguin (DP) contributions to
$\DMds$ and $\BRBds$ has been extensively discussed in the literature
\cite{BCRS, Isidori:2001fv, D'Ambrosio:2002ex, Carena:2006ai, recent,
  other}. In the following, we extend these analyses by incorporating
the effects due to the new term $\epsGP$. This allows us to obtain
improved predictions for $\DMds$ and $\BRBds$ in the large $\tb$
regime of the MFV MSSM based on the $SU(2) \times U(1)$ symmetry
limit.

Predicting the $\BdsBbards$ mass differences involves integrating out
heavy degrees of freedom at a scale of order $\mt$ by matching on to 
the effective Hamiltonian   
\beq \label{eq:HeffDB2}
\Heff^{\Delta B = 2} = \frac{\GF^2 \MW^2}{16 \pi^2} (V_{tb}^{{\rm eff}
  \ast} V_{td^i}^{\rm eff})^2 \sum_j C_j Q_j + \hc
\eeq

In the MFV MSSM with large $\tb$ the numerically dominant contributions
to $\DMds$ are induced by the two effective operators   
\beq \label{eq:Q2Q1}
Q_2^{\rm LR} = (\bar{b}_{\sR} d^i_{\sL}) (\bar{b}_{\sL} d^i_{\sR}) \,
,\hspace{2.5mm} Q_1^{\rm SLL} = (\bar{b}_{\sR} d^i_{\sL})
(\bar{b}_{\sR} d^i_{\sL}) \, .
\eeq

Combining the flavor-changing neutral Higgs couplings of
\Eq{eq:lagrangian} we find that the initial conditions of the
corresponding Wilson coefficients are given by
\beq \label{eq:C2C1}
\begin{aligned} 
  C_2^{\rm LR} & = -\frac{\GF \mbbar \mdibar \mtbar^4}{\sqrt{2} \pi^2
    \MW^2} \frac{(16 \pi^2)^2 \, \TB^4 \, \epsY^2 \, \cal{F}^+}{(1 +
    \epsthree \, \TB)^2 (1 + \epszero \, \TB)^2} \, , \hspace{2mm} \\
  C_1^{\rm SLL} & = -\frac{\GF \mbbar^2 \mtbar^4}{2 \sqrt{2} \pi^2
    \MW^2} \frac{(16 \pi^2)^2 \, \TB^4 \, \epsY^2 \, \cal{F}^-}{(1 +
    \epsthree \, \TB)^2 (1 + \epszero \, \TB)^2} \, ,
\end{aligned}
\eeq
where 
\beq \label{eq:Fpm} 
\begin{split}
  {\cal F}^\pm & = \frac{\SAB^2}{\MH^2} + \frac{\CAB^2}{\mh^2} \pm
  \frac{1}{\MA^2} \\
  & + \frac{\CA \, ( 2 \CAB + \CA \, \deltaGP) \, \deltaGP}{\mh^2} \,
  ,
\end{split}
\eeq
and 
\beq \label{eq:dGP}
\deltaGP = \frac{\mh^2}{\MH^2 - \mh^2} \, \epsGP \, .
\eeq
Here we have used the approximations $\sin \beta \simeq 1$ and $\cos
\beta \simeq 0$ valid for $\tb \gg 1$. The same relations are employed
in the following whenever it is justified.  The first line of
\Eq{eq:Fpm} resembles the result derived first by the authors of
\cite{BCRS}, while the second one represents the new contribution to
the factors $\cal{F}^\pm$ due to $\epsGP$.

Using $\DMds = | \langle B_{d,s} | \Heff^{\Delta B = 2} |
\bar{B}_{d,s} \rangle |/M_{B_{d,s}}$ one obtains from \Eq{eq:HeffDB2}
the DP contribution to the $\BdsBbards$ mass differences. To an
excellent approximation one has      
\beq \label{eq:DMdiDP}
\begin{split}
\Delta M_{d^i}^{\rm DP} & = \f{\GF^2 \MW^2}{24 \pi^2} M_{B_{d^i}}
f_{B_{d^i}}^2 \times \\  
& |V_{tb}^{{\rm eff} \ast} V_{td^i}^{\rm eff}|^2 \left ( P_2^{\rm LR} 
C_2^{\rm LR} + P_1^{\rm SLL} C_1^{\rm SLL} \right ) \, ,  
\end{split}
\eeq 
where the factors $P_2^{\rm LR} = 2.56$ and $P_1^{\rm SLL} = -1.06$
condense renormalization-group-improved NLO QCD corrections
\cite{Buras:2000if} and the relevant matrix elements
\cite{Becirevic:2001xt}. In our numerical analysis we employ the
unquenched staggered three flavor results $f_{B_d} = 216 (22)
\hspace{0.25mm} \MeV$ \cite{Gray:2005ad} and $f_{B_s} = 260 (29)
\hspace{0.25mm} \MeV$ \cite{Wingate:2003gm} for the $B_{d,s}$-meson
decay constants obtained by the HPQCD Collaboration, while we take $|
V_{tb}^{{\rm eff} \ast} V_{td}^{\rm eff}| = 86 (14) \times 10^{-4}$,
$| V_{tb}^{{\rm eff} \ast} V_{ts}^{\rm eff}| = 41.3 (7) \times
10^{-3}$ \cite{Ball:2006xx}, $M_{B_d} = 5.2793 \hspace{0.25mm} \GeV$,
and $M_{B_s} = 5.3696 \hspace{0.25mm} \GeV$ \cite{Yao:2006px}.
             
In the case of the rare decays $\Bdsmm$, the effective Hamiltonian 
that arises after removing all heavy particles as active degrees of
freedom is given by 
\beq \label{eq:HeffBll}
\Heff^{\Delta B = 1} = -\frac{\GF}{\sqrt{2}} \frac{\aem}{\pi \sws}
V_{tb}^{{\rm eff} \ast} V_{td^i}^{\rm eff} \sum_j C_j Q_j + \hc \, ,
\eeq
where the electromagnetic coupling $\aem$ and the weak mixing angle
$\sw$ are naturally evaluated at the electroweak scale
\cite{Bobeth:2003at}.  

In the large $\tb$ regime of the MFV MSSM only two effective operators
can have a sizable impact on $\Bdsmm$, namely  
\beq \label{eq:QSQP}
Q_S = \mb (\bar{b}_{\sR} d^i_{\sL}) (\bar{\mu} \mu) \, ,
\hspace{2.5mm} Q_P = m_{d^i} (\bar{b}_{\sL} d^i_{\sR}) (\bar{\mu}
\gamma_5 \mu) \, .
\eeq

The same flavor-changing Higgs vertices that generate the dominant
contribution to $\DMds$ induce enhanced $\tb$ corrections to the
Wilson coefficients of the semileptonic operators $Q_S$ and $Q_P$. The
effective couplings of \Eq{eq:lagrangian} lead to the following
matching conditions
\beq \label{eq:CSCP}
\begin{split}
  & C_S = \frac{\mmu \mtbar^2}{4 \MW^2} \frac{16 \pi^2 \,\TB^3 \,
    \epsY}{(1 + \epsthree \, \TB) (1 + \epszero \, \TB)} \Bigg [
  \frac{\CA
    \SAB}{\MH^2} \\
  & -\frac{\SA \CAB}{\mh^2} - \frac{\CA \, ( \CAB + \SA + \CA \,
    \deltaGP) \, \deltaGP}{\mh^2} \Bigg ] \, , \\
  & C_P = -\frac{\mmu \mtbar^2}{4 \MW^2} \frac{16 \pi^2 \,\TB^3 \,
    \epsY}{(1 + \epsthree \, \TB) (1 + \epszero \, \TB)}
  \frac{1}{\MA^2} \, ,
\end{split}
\eeq
where $\mmu = 105.66 \hspace{0.5mm} \MeV$ \cite{Yao:2006px}. While
$C_P$ remains unchanged with respect to the analytic expression first
presented in \cite{BCRS}, $C_S$ picks up an additional contribution
due to $\epsGP$ given by the second term in the second line of
\Eq{eq:CSCP}.

The DP contribution to the branching ratios of $\Bdsmm$ can be
expressed to a very good approximation in terms of the initial
conditions $C_S$ and $C_P$ as 
\beq \label{eq:BRBdimm}
\begin{split}
  \BR (B_{d^i} \to \mu^+ \mu^-)^{\rm DP} & = \f{\GF^2 \aem
    M_{B_{d^i}}^5
    f_{B_{d^i}}^2 \tau_{B_{d^i}}}{64 \pi^3} \times \\
  & |V_{tb}^{{\rm eff} \ast} V_{td^i}^{\rm eff}|^2 \left ( | C_S |^2 +
    | C_P |^2 \right ) \, ,
\end{split}
\eeq
where the $B_{d,s}$-meson lifetimes are taken to be $\tau_{B_d} =
1.527 \hspace{0.25mm} {\rm ps}$ and $\tau_{B_s} = 1.454
\hspace{0.25mm} {\rm ps}$ \cite{Barberio:2006bi}.

\section{Numerical analysis} 
\label{sec:numerics}

We are now in the position to analyze the impact of the correction
$\epsGP$ on the prediction of $\DMds$ in the MFV MSSM with large
$\tb$, taking into account the constraints from the low-energy
observables $\DMK$, $\epsK$, $\BRga$, $\BRll$, $\BRBds$, and $\BRBtn$,
as well as the limit on the lightest Higgs boson mass $\mh$. In the
calculation of the flavor physics observables all relevant
contributions stemming from $A^0$, $h^0$, $H^0$, $H^\pm$, and
$\tilde{\chi}^\pm$ exchange are taken into account. More precisely, in
the case of $\DMK$ and $\epsK$ we rely on the formulas given in
\cite{Carena:2006ai}, while our calculation of $\BRga$ includes all
$\tb$ enhanced charged Higgs and chargino corrections \cite{ltbup} as
well as the DP contribution \cite{D'Ambrosio:2002ex}. In the case of
$\DMs$ we combine the neutral Higgs effects with the $\tb$ resummed
terms from charged Higgs box diagrams \cite{BCRS, Isidori:2001fv},
while for $\BRBtn$ we employ the formula first derived in
\cite{Akeroyd:2003zr}. In all cases we supplement the existing
expressions with the corrections stemming from the new term $\epsGP$
and use the complete formulas \Eqsto{eq:epstilde3}{eq:epsYprime} for
$\epsthree$, $\epszero$, $\epsY$, $\epszero^\prime$, and
$\epsY^\prime$.

Another important difference with respect to the preceding analyses
\cite{BCRS, Isidori:2001fv, D'Ambrosio:2002ex, Carena:2006ai, recent}
is the fact that we do not evaluate the mixing angle $\alpha$ and the
masses $\mh$ and $\MH$ appearing in \Eqsand{eq:Fpm}{eq:CSCP} at
tree-level, but include the dominant one-loop corrections
\cite{mhsusy}, which are essential to obtain $\mh > \MZ$. While the
inclusion of these higher order corrections turns out to have a minor
impact on $\DMs$ and $\Bdsmm$, we find that they have a profound
effect on $\DMd$, as they invalidate the common assumption \cite{BCRS,
  Isidori:2001fv, D'Ambrosio:2002ex, Carena:2006ai, recent} that
$C_1^{\rm SLL}$ gives only a negligible contribution to the $\BdBbard$
mass difference.\footnote{In \cite{Parry:2006vq} it has been claimed
  that in $\DMs$ the contribution due to $C_1^{\rm SLL}$ may amount to
  $80 \% \, (45 \%)$ of $C_2^{\rm LR}$ for $\MA = 150 \hspace{0.5mm}
  \GeV \, (200 \hspace{0.5mm} \GeV)$. We are unable to reproduce these
  results.}

\begin{figure}[!t]
\begin{center}
\hspace{-5mm}% 
\scalebox{0.5}{\rotatebox{270}{\includegraphics{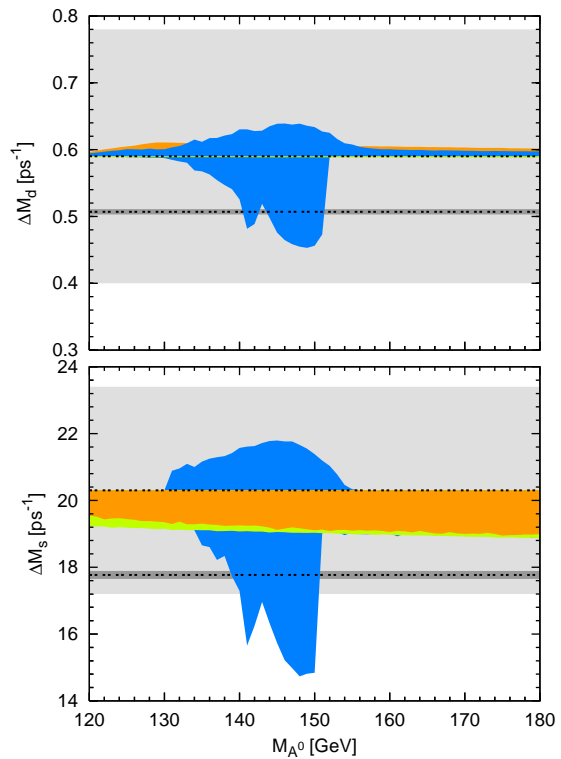}}}
\end{center}
\vspace{-5mm}
\caption{$\DMd$ ($\DMs$) as a function of $\MA$. The blue (dark gray)
  areas correspond to the improved predictions, while the orange
  (medium gray) and yellow-green (light gray) regions are obtained by
  neglecting sequentially the contributions due to $\epsGP$ and the
  one-loop corrections to $\alpha$, $\mh$, and $\MH$. The $1 \mysigma$
  ranges and central values of the experimental (SM) results are
  indicated by the dark (light) gray bands underlying the dotted
  lines. See text for details.}
\label{fig:DMds}
\end{figure}

In our numerical analysis we focus on scenarios with heavy sparticles
and the mass scale of the Higgs sector close to the electroweak
scale. We allow the parameters to float freely in the following
ranges: $10 \leq \tb \leq 60$, $100 \hspace{0.5mm} \GeV \leq \MA \leq
500 \hspace{0.5mm} \GeV$, $1 \hspace{0.5mm} \TeV \leq \tilde{m}_i \leq
2 \hspace{0.5mm} \TeV$ where $\tilde{m}_{i} = m_{\tilde{t}_L},
m_{\tilde{t}_R}, m_{\tilde{b}_R}, m_{\tilde{\tau}_L},
m_{\tilde{\tau}_R}, M_1, M_2, \mgl$, and $1 \hspace{0.5mm} \TeV \leq |
\tilde{M}_i | \leq 2 \hspace{0.5mm} \TeV$ for $\tilde{M}_i = \mu, A_t,
A_b, A_\tau$. The SM parameters are fixed to $\as = \asMSbar (\mtbar)
= 0.109$, $\mtbar = 165 \ \GeV$, $\mbbar = 3\ \GeV$, $\msbar = 0.06 \
\GeV$, $\mdbar = 0.003 \ \GeV$, $\mlbar = 1.78 \ \GeV$, $\aem =
\aemMSbar (\MZ) = 1/127.9$, and $\sws = \sin^2 \hat{\theta}_W = 0.231$
\cite{Yao:2006px}. In order to find the boundaries of the allowed
parameter space we perform an adaptive scan of the 14 SUSY variables
employing the method advocated in \cite{adaptive}.  The correctness of
the obtained results has been independently verified by a scanning
procedure based on random walk techniques.

To simplify our numerical analysis we set all CKM factors and
non-perturbative parameters to their central values and combine
experimental and theoretical uncertainties into bounds corresponding
to $95 \%$ confidence levels (CLs) by adding theory errors
linearly. Severe constraints on the SUSY parameter space follow from
$\BRga$, $\BRBs$, $\BRBtn$, and $\mh$. In the case of $\BXsga$ the
most recent SM calculations \cite{bsg} are used and $\BRga$ is
required to lie in the interval $2.7 \leq \BRga \times 10^4 \leq
4.4$. Since the SM prediction of $\BRga$ is now lower than the
experimental world average \cite{Barberio:2006bi} by about $1.4
\mysigma$, a cancellation between the constructive charged Higgs
corrections and the chargino contribution is easier to achieve than in
the past, where the theoretical result used to be above the
experimental one. As far as $\Bsmm$ is concerned all parameter points
are required to satisfy $\BRBs < 1.0 \times 10^{-7}$
\cite{Bernhard:2006fa}, while, in view of the sizable experimental
\cite{Btaunu} and theoretical uncertainties, we use $0.2 \leq \BRBtn
\times 10^4 \leq 2.5$ in the case of $\Btn$. Because the interference
between SM and charged Higgs corrections is necessarily destructive
\cite{Hou:1992sy}, $\Btn$ may become the most stringent constraint in
the near future, in particular if improved measurements of $\BRBtn$
will not differ much from the SM expectation. Finding $\BRBtn$ close
to its SM prediction would also have a important effect on our
numerical analysis since light values of $\MA$ would be disfavored in
such a case. Concerning the lightest neutral Higgs boson we ensure
$\mh > 114.4 \hspace{0.5mm} \GeV$ \cite{ewpm}.

The constraints from $K$- and the remaining $B$-physics observables
are much less restrictive. $\DMK$ and $\epsK$ are allowed to differ
from their experimental values \cite{Yao:2006px} by $\pm 50 \%$ and
$\pm 40 \%$, while we reject parameter points that reverse the sign of
the amplitude ${\cal A} (\btosgamma)$ with respect to the SM, as they
correspond to $\BRll$ values higher than the measurements \cite{bxsll}
by around $3 \mysigma$ \cite{Gambino:2004mv}. In the case of $\Bdmm$
we require $\BRBd < 3.0 \times 10^{-8}$ \cite{Bernhard:2006fa}. Notice
that, we do not take into account the experimental constraint from $(g
- 2)_\mu$, because in our scenario the slepton sector parameters are
uncorrelated with the ones of the squark sector, so that $(g - 2)_\mu$
does not lead to any restriction.

The MFV MSSM prediction for the mass differences $\DMds$ as a function
of $\MA$ can be seen in \Fig{fig:DMds}. The blue (dark gray) areas
correspond to the full results obtained from
\Eqsto{eq:C2C1}{eq:DMdiDP}, while the orange (medium gray) and
yellow-green (light gray) regions are obtained after removing
successively the contributions due to $\epsGP$ and the one-loop
corrections to $\alpha$, $\mh$, and $\MH$. For comparison, the $68 \%
\ {\rm CLs}$ and central values of the measurements $\DMd^{\rm exp} =
0.507 (4) \hspace{0.5mm} {\rm ps}^{-1}$ \cite{Barberio:2006bi} and
$\DMs^{\rm exp} = 17.77 (12) \hspace{0.5mm} {\rm ps}^{-1}$
\cite{Abulencia:2006ze} and the SM expectations $\DMd^{\rm SM} = 0.59
(19) \hspace{0.5mm} {\rm ps}^{-1}$ and $\DMs^{\rm SM} = 20.3 (3.1)
\hspace{0.5mm} {\rm ps}^{-1}$ \cite{Dalgic:2006gp} are indicated by
the dark and light gray bands underlying the dotted lines. The
prediction $\DMd^{\rm SM}$ is obtained from the central value of
$\DMs^{\rm SM}$ using $\xi = f_{B_s} \hat{B}_{B_s}^{1/2}/(f_{B_d}
\hat{B}_{B_d}^{1/2}) = 1.216 (41)$ \cite{Okamoto:2005zg} and the CKM
factors and $B_{d,s}$-meson masses given earlier by adding all errors
in quadrature. For a critical discussion of hadronic uncertainties in
$\DMds^{\rm SM}$ we refer to \cite{Ball:2006xx}.

From \Fig{fig:DMds} it is evident that, whereas one-loop corrections
to $\alpha$, $\mh$, and $\MH$ have only a minor impact on $\DMs$,
which slowly loses importance with increasing $\MA$, they are
essential to obtain a correct prediction in the case of $\DMd$. While
already this result is interesting by itself, truly spectacular
effects arise after the inclusion of the new term $\epsGP$. Now large
negative and positive corrections to $\DMd$ ($\DMs$) of up to $-0.14
\hspace{0.5mm} {\rm ps}^{-1}$ and $0.05 \hspace{0.5mm} {\rm ps}^{-1}$
($-5.6 \hspace{0.5mm} {\rm ps}^{-1}$ and $1.5 \hspace{0.5mm} {\rm
  ps}^{-1}$) are possible in the mass window $130 \hspace{0.5mm} \GeV
\lesssim \MA \lesssim 160 \hspace{0.5mm} \GeV$ without violating any
existing constraint from flavor and collider physics. These large
corrections typically arise for $15 \lesssim \tb \lesssim 30$ and
$\mu, A_t \gtrsim 1.5 \hspace{0.5mm} \GeV$. Their size is (slightly)
less pronounced for larger (smaller) values of $\mstl$ and $\mstr$
($\mgl$), while they are highly uncorrelated with the remaining SUSY
parameters. Therefore they do not correspond to exceptional points in
the large $\tb$ and small $\MA$ region of the MSSM parameter space.

As pointed out above, the small $\MA$ region may be severly constrained by
a more precise determination of $\BRBtn$. The large effects shown in
\Fig{fig:DMds} occur only for $\BRBtn \lesssim 10^{-4}$. If future measurements
should find a value in the ballpark of $1.5 \times 10^-4$ 
with a small error of 30\% or less, 
large corrections to $\DMds$ would be excluded within the MSSM with MFV.

\begin{figure}[!t]
\begin{center}
\hspace{-5mm}% 
\scalebox{0.5}{\rotatebox{270}{\includegraphics{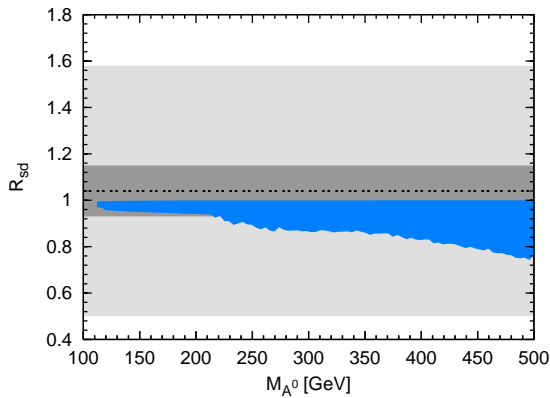}}}
\end{center}
\vspace{-5mm}
\caption{$R_{sd}^{\rm DP}$ as a function of $\MA$. The blue (dark
  gray) area corresponds to the improved MFV MSSM prediction. The $1
  \mysigma$ range and central value of the present (a possible future)
  tree-level prediction $R_{sd}^{\rm exp}$ corresponding to $\gamma =
  67(31)^\circ$ ($\gamma = 67(5)^\circ)$ is indicated by the light
  (dark) gray band and the dotted horizontal line. See text for
  details.}
\label{fig:Rsd}
\end{figure}

It is also compelling to analyze the impact of neutral Higgs DP
contributions in the double ratio $R_{sd}^{\rm DP} = (\DMs^{\rm
  DP}/\DMs^{\rm SM})/(\DMd^{\rm DP}/\DMd^{\rm SM})$, which has smaller
hadronic uncertainties than the ratios $\DMds^{\rm DP}/\DMds^{\rm SM}$
themselves. The size of the possible departures of $R_{sd}^{\rm DP}$
from the SM value $1$ should be compared to the total uncertainty of
the double ratio 
\bea \label{eq:Rsdexp}
\begin{split}
  R_{sd}^{\rm exp} & = \lambda^2 \left ( 1 - 2 R_b \cos \gamma + R_b^2
  \right ) \times \\
  & \left ( 1 + \left ( 1 - 2 R_b \cos \gamma \right ) \lambda^2
  \right ) \f{1}{\xi^2} \f{M_{B_d}}{M_{B_s}} \f{\DMs^{\rm
      exp}}{\DMd^{\rm exp}} \, , \hspace{5mm}
\end{split}
\eea that can be determined almost independently of new physics and $R_b
= (1 - \lambda^2/2) \hspace{0.5mm} 1/\lambda \hspace{0.5mm}
|V_{ub}^{\rm eff}/V_{cb}^{\rm eff}|$ \cite{Ball:2006xx} from $\lambda
= |V_{us}^{\rm eff}|$, $M_{B_{d, s}}$, $\DMds^{\rm exp}$, $\xi$ and
the reference unitarity triangle angle $\gamma$ measured in tree-level
dominated $B$-decays like $B \to D^{(\ast)} K^{(\ast)}$.

The double ratio $R_{sd}^{\rm DP}$ as a function of $M_{A^0}$ is shown
in \Fig{fig:Rsd}, where the blue (dark gray) area represents the full
result derived from \Eqsto{eq:C2C1}{eq:DMdiDP}. The $68 \% \ {\rm CL}$
and central value of the tree-level determination $R_{sd}^{\rm exp} =
1.04(54)$ is indicated by the light gray band and the dotted line. The
quoted range of $R_{sd}^{\rm exp}$ is obtained from \Eq{eq:Rsdexp}
using $\lambda = 0.2257(21)$, $R_b = 0.42(4)$, $\gamma = 67(31)^\circ$
\cite{Charles:2004jd} and the remaining input specified above by
adding all uncertainties in quadrature. Because of the poor knowledge
of $\gamma$ from $B \to D^{(\ast)} K^{(\ast)}$ the double ratio
$R_{sd}^{\rm exp}$ is only weakly constrained at the moment.

Two properties of $R_{sd}^{\rm DP}$ deserve a special mention. Our
numerical analysis reveals that $i)$ effects due to $\eps_{\rm GP}$
and the one-loop corrections to $\alpha$, $\mh$, and $\MH$ cancel
almost entirely in the double ratio $R_{sd}^{\rm DP}$ since their size
is strongly correlated between $\DMd^{\rm DP}$ and $\DMs^{\rm DP}$ and
$ii)$ the MFV MSSM with large $\tb$ predicts $R_{sd}^{\rm DP} \leq 1$
as demonstrated in \Fig{fig:Rsd}. The impact of a future precision
measurement of $\gamma$ by the LHCb experiment \cite{LHCb} is also
illustrated in this figure. Assuming $\gamma = 67(5)^\circ$ but $\xi$
unaltered leads to $R_{sd}^{\rm exp} = 1.04(11)$. The corresponding
$68 \% \ {\rm CL}$ and central value is indicated by the dark gray
band and the dotted line in \Fig{fig:Rsd}. It can be seen that even
with this improvement in precision, $R_{sd}^{\rm exp}$ offers only
limited potential for exclusion of new physics through a deviation
from its SM value. A similar conclusion has been drawn in
\cite{Ball:2006xx}.

\section{Conclusions} 
\label{sec:conclusions}

To conclude, we have pointed out that in the MSSM terms from the
mixing between eigenstates of the two Higgs doublets give rise to
$\tb$ enhanced corrections that do not vanish in the limit where the
SUSY particles are much heavier than the Higgs bosons. We have
calculate all one-loop corrections of this type by matching the full
MSSM on to an effective two-Higgs-doublet model of type II. After
combining the missing effective couplings with all known
non-holomorphic terms we obtain improved predictions for the
$\BdsBbards$ mass differences $\DMds$ and the branching ratios of
$\Bdsmm$ in the large $\tb$ regime of the MSSM with
minimal-flavor-violation. Our numerical analysis shows that these
universal contributions have striking phenomenological implications as
they can lead to large SUSY effects in $\DMd$ and to a significant
enhancement of $\DMs$ relative to the SM prediction for a light
pseudoscalar Higgs boson $A^0$ with mass in the range between $130
\hspace{0.5mm} \GeV$ and $160 \hspace{0.5mm} \GeV$.

\acknowledgments{We are indebted to U.~Nierste for suggesting the
  topic and to M.~Gorbahn for sharing related unpublished work with
  us. Very helpful correspondence with S.~J\"ager and D.~St\"ockinger is
  acknowledged. Special thanks goes to A.~J.~Buras, M.~Gorbahn,
  G.~Isidori, U.~Nierste, and P.~Slavich for a careful reading of the manuscript
  and for all their critical comments. This work has been supported in
  part by the Schweizer Nationalfonds.}

\end{document}